\documentclass[11pt,twoside]{article}

\usepackage{asp2006}
\usepackage{epsf}
\usepackage{graphicx}
\usepackage{lscape}

\def\la{\mathrel{\mathchoice {\vcenter{\offinterlineskip\halign{\hfil
$\displaystyle##$\hfil\cr<\cr\sim\cr}}}
{\vcenter{\offinterlineskip\halign{\hfil$\textstyle##$\hfil\cr
<\cr\sim\cr}}}
{\vcenter{\offinterlineskip\halign{\hfil$\scriptstyle##$\hfil\cr
<\cr\sim\cr}}}
{\vcenter{\offinterlineskip\halign{\hfil$\scriptscriptstyle##$\hfil\cr
<\cr\sim\cr}}}}}

\def\ga{\mathrel{\mathchoice {\vcenter{\offinterlineskip\halign{\hfil
$\displaystyle##$\hfil\cr>\cr\sim\cr}}}
{\vcenter{\offinterlineskip\halign{\hfil$\textstyle##$\hfil\cr
>\cr\sim\cr}}}
{\vcenter{\offinterlineskip\halign{\hfil$\scriptstyle##$\hfil\cr
>\cr\sim\cr}}}
{\vcenter{\offinterlineskip\halign{\hfil$\scriptscriptstyle##$\hfil\cr
>\cr\sim\cr}}}}}

\begin{document}
\title{Stellar wind accretion in high-mass X-ray binaries}   
\author{I. Negueruela}   
\affil{DFISTS, EPSA, Universidad de Alicante, Apdo. 99,
  03080 Alicante, Spain}    

\begin{abstract} 

Recent discoveries have confirmed the existence of a large population
of X-ray sources fuelled by accretion from the stellar wind of an OB
supergiant. Such systems are powerful laboratories to study many
aspects of astrophysics. Over the last decades, the physics of
accretion in these systems has been the subject of extensive research,
mainly through numerical methods. In spite of this effort, large
uncertainties remain in our understanding, reflecting the complexity
of the physical situation. A crucial issue that remains open is the
possible formation of accretion discs. Though the spin evolution of
neutron stars in these systems suggests that angular momentum is, at least
occasionally, accreted, and many observational facts seem to require
the existence of discs, computational results do not favour this
possibility. In this brief review, I will summarise some of the open
questions in this area.

\end{abstract}


\section{Introduction}

 High-Mass X-ray binaries (HMXBs) are X-ray sources composed of an
early-type massive star and an accreting compact object (neutron star
or black hole). The vast majority of HMXBs contain X-ray pulsars (magnetised
neutron stars) and can be characterised by their pulse period, which
corresponds to the spin period of the neutron star
$P_{{\rm s}}$. A few systems are considered good black
hole candidates. The 
presence of an X-ray pulsar has allowed the
determination of orbital parameters for a large number of systems via
analysis of Doppler shifts in the pulse arrival times
(e.g., Schreier et al. 1972; Rappaport et al. 1978). 

Since so many
parameters may be known, HMXBs represent important laboratories to
study a large number of fundamental astrophysical questions. Among
them, we can cite the masses of neutron stars and their equation of
state (e.g., Kaper 1998; Quaintrell et al. 2003; van der Meer
2007). Also, because of their young age, they may act as tracers of
star formation (e.g., Grimm et al. 2003; Lutovinov et al. 2005). When
considered as a    
  population, they can provide information on properties of galaxies
  (e.g., Gilfanov et al. 2004). Finally, they 
  represent an important phase of massive binary 
  evolution and, again considered as a whole, can provide strong
  constraints on binary evolution and the mechanisms for the formation
  of neutron stars and black holes (e.g., van Bever \& Vanbeveren
  2000; Brown et al. 2001). A recent list of known
  HMXBs is provided by Liu et al. (2007).

HMXBs can be divided in several subgroups, depending on their X-ray
properties, which are found to depend fundamentally on the nature of
the massive star donating mass. Corbet (1986) found that the position
of an object in the orbital period vs. spin period ($P_{{\rm 
    orb}}$/$P_{{\rm s}}$) diagram correlates well with other
physical properties, allowing the definition of meaningful
subgroups:

\begin{itemize}

\item A few objects with very close orbits and short $P_{{\rm s}}$
are observed as very bright
     ($L_{{\rm X}}\sim 10^{38}\:\mathrm{erg\,s^{-1}}$) persistent X-ray
    sources, with clear evidence for an accretion disc. Their neutron
    stars are spinning up most of the time, because 
    angular momentum is transferred with the material accreted. 

\item Many systems
    with Be star counterparts are  X-ray transients, showing very low
X-ray fluxes in quiescence ($\la10^{33}\:\mathrm{erg\,s^{-1}}$; e.g., Campana et
al. 2002) and luminosities close to the Eddington limit for a
neutron star ($\sim10^{38}\:\mathrm{erg\,s^{-1}}$) during bright outbursts. In
spite of their short duty cycles, the majority of known HMXBs are
Be/X-ray transients, suggesting that they represent the dominant
population. Their X-ray characteristics are explained as a consequence
of mass loss from the Be star in the form of a Keplerian disc (Okazaki
\& Negueruela 2001). Be/X-ray binaries populate a rather narrow
region of the  $P_{{\rm orb}}$/$P_{{\rm s}}$ diagram. The correlation
between $P_{{\rm orb}}$ and $P_{{\rm s}}$ is believed to be connected
to some physical equilibrium between the spin down of the neutron
stars when they are not accreting and their spin up during outbursts,
indicative of the formation of transitory accretion discs (Stella et
al. 1986; Waters \&
van Kerkwijk 1989; Wilson et al. 2008).

\item A second large group is formed by systems with OB supergiant
  donors and long-$P_{{\rm s}}$ neutron stars. These objects, known as
  Supergiant X-ray Binaries (SGXBs) are
  moderately bright ($L_{{\rm X}}\sim 10^{36}\:\mathrm{erg\,s^{-1}}$) 
  persistent X-ray sources with a large degree of short-term
  variability. Their characteristics are attributed to direct accretion from the
  wind of the supergiant on to the compact object. This mode of
  accretion is known as wind accretion, and, as we will discuss, it is
  believed to occur without the formation of a stable
  accretion disc.

\end{itemize}

Recent discoveries, however, are showing that the division of HMXBs
into different groups, though physically meaningful, is not
strict. Apparent exceptions have been found, which may perhaps be due
to infrequent evolutionary stages (cf.
Chaty, these proceedings).

\section{The theory of wind accretion}

Luminous OB stars produce powerful winds due to line scattering of the
continuum radiation flux from the star (see Kudritzki \& Puls 2000,
for a review). Material is
accelerated outwards from the stellar atmosphere to a final velocity
$v_{\infty}$ according to a law that may be approximated as
\begin{equation}
v_{\rm w}(r)=v_{\infty}\left(1-\frac{R_{*}}{r}\right)^\beta
\, ,
\end{equation}
where $R_{*}$ is the radius of the OB star and $\beta$ is a factor
generally lying in the interval $\sim0.8-1.2$. 

In these circumstances, high wind velocities are reached at a moderate
height above the star's surface (for example, for $\beta=1.0$, $v_{\rm
  w}(2R_{*})=0.5v_{\infty}$, i.e., $\sim1000\;{\rm km}\,{\rm
  s}^{-1}$). The wind is highly supersonic and the wind speed is much
higher than the orbital velocity of the neutron star $v_{{\rm
    orb}}\sim200\;{\rm km}\,{\rm 
  s}^{-1}$ at $2\,R_{*}$). Under those conditions, the classical
Bondi-Hoyle-Lyttleton formulation represents an approximation to the accretion
process. A comprehensive review of Bondi-Hoyle-Lyttleton accretion is presented
by Edgar (2004).

In this approximation, the accretion rate ($\dot{M}$) of a neutron
star immersed in a fast
wind is governed by its accretion radius, the maximum distance at which its
gravitational potential well can deflect the stellar wind and focus
the outflowing material towards the neutron star, given by
\begin{equation}
 r_{{\rm acc}} \sim \frac{2GM_{{\rm X}}}{v^{2}_{{\rm rel}}}
\, ,
\end{equation}
where the relative velocity of the accreted
material with respect to the neutron star is $v^{2}_{{\rm
    rel}}= v^{2}_{\rm w} + v^{2}_{{\rm orb}}$. For typical values of
$v_{\rm w}$,  $r_{{\rm acc}}\sim 10^{8}\:{\rm m}$, decreasing by a factor of a
few from $r = 2\:R_{*}$ to $r = 10\:R_{*}$. 

The accretion rate is then
\begin{equation}
\dot{M}=4\pi\left( GM_{{\rm X}}\right)^{2}\frac{v_{\rm
    w}(r)\rho(r)}{v^{2}_{{\rm rel}}}\, ,
\end{equation}
where $\rho(r)$ is the wind density. If we assume a certain efficiency
factor in the transformation of gravitational energy into X-ray
luminosity, the X-ray luminosity will be proportional to 
\begin{equation}
L_{{\rm X}}\sim \dot{M} \sim \frac{\rho(r)v_{\rm w}(r)}{v^{4}_{{\rm rel}}}
\, .
\end{equation}

This dependency means that, in this approximation, the X-ray
luminosity of a neutron star in a circular orbit is constant. In an
eccentric orbit, the luminosity varies with orbital phase. Different
systems differ in their orbital parameters (semi-major axis,
eccentricity) and wind properties (mass loss rate, velocity law and
terminal velocity). Figure~1 is a simple diagram showing the
dependence of the X-ray luminosity on these parameters. A
transformation factor  $L_{\rm X}=\frac{1}{3}L_{\rm acc}$ is assumed.

\begin{figure*}[ht!]
\label{fig:lx}
   \centering
   \resizebox{\textwidth}{!}{\includegraphics{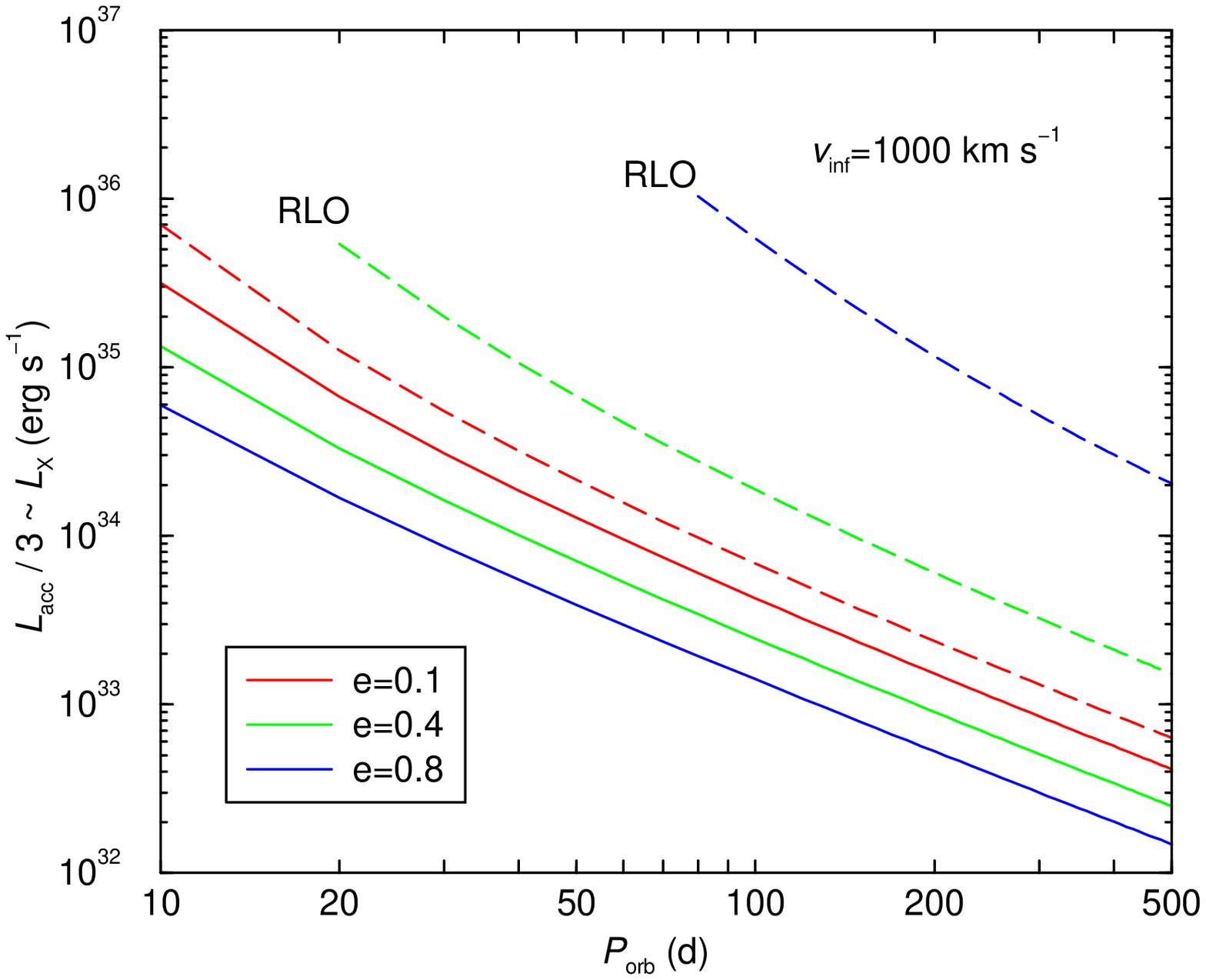}\includegraphics{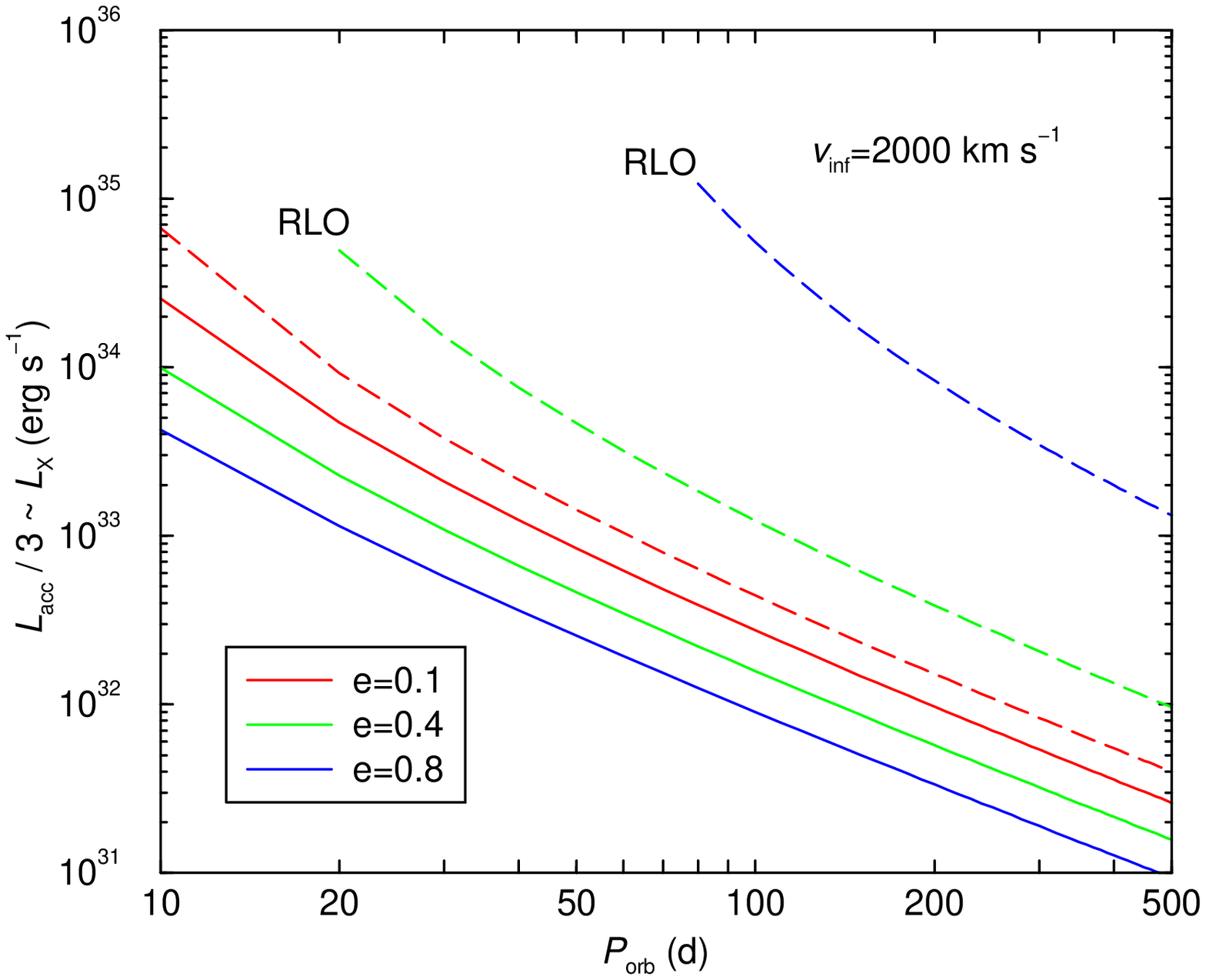}}
\caption{Minimum and maximum X-ray luminosities (within an orbit)
  predicted by a simple approximation (homogeneous wind +
  Bondi-Hoyle-Lyttleton 
  accretion) for a range of system parameters ($P_{{\rm orb}}$,
  $e$). The left panel shows 
  luminosities for a {\it slow} wind ($v_{\infty}=1000\:{\rm km}\,{\rm
  s}^{-1}$), typical of B-type supergiants. The right panel shows
  results for a {\it fast} wind  ($v_{\infty}=2000\:{\rm km}\,{\rm
  s}^{-1}$), more typical of O-type supergiants. The regions marked
  RLO are those where the donor undergoes Roche Lobe Overflow close to
periastron. Courtesy of M.~Rib\'o.}
\end{figure*}

\section{Real winds and real accretion}
Direct application of the simple model outlined above is able to
crudely reproduce the average order of magnitude X-ray luminosity of
real SGXBs \citep{ribo06}. However, their X-ray lightcurves provide
strong evidence for a much more complex physical situation. The X-ray
lightcurves of wind accreting systems are dominated by short-term
(timescale of hundreds of seconds) variability, with flux changes by a
factor of a few \citep[e.g.,][]{haberl89,krey99}. The pulse  
periods of well-monitored 
SGXBs, such as Vela X-1, display random variations around an average
value \citep{bild97}. These variations can occur on short (days)
timescales \citep{boynton84}, 
superimposed on longer-timescale trends
\citep{nagase84}. The behaviour has been described as a random walk
\citep[e.g.,][]{tsu89}, and interpreted as short episodes of angular
momentum accretion.

Some systems have shown very long term trends. For example,
4U~1907+09 showed persistent spin down, at an almost constant rate,
during more than ten years. Afterwards, it changed to a lower
spin-down rate and finally started spinning up \citep{fritz06}. These
``torque reversals'' have been taken as evidence for accretion discs
that rotate in alternating senses \citep{nelson97}. Interestingly,
these torque reversals are not only seen in wind-fed systems, but also
in systems that always accrete from a disc, such as Cen X-3 \citep{nelson97}.

In order to explain this behaviour, the complex physics of SGXBs must
be considered.
In short, we may identify three main ways in which the actual physical
process of accretion differs from the idealised formulation of the
previous section:
\begin{enumerate}
\item The winds of massive stars are highly structured.
\item Accretion on to a very small object is an unstable process.
\item The magnetic field of the neutron star may affect the flow of
  material.
\end{enumerate}

\subsection{Complex radiative winds}

 From the early days of
radiation driven wind theory \citep[e.g.,][]{or84}, it was obvious that 
winds from hot stars should be highly unstable to small-scale
perturbations. This is a direct consequence of the driving mechanism
and therefore an intrinsic property of the wind, which cannot be
smooth, but must be highly structured. Observations of wind lines in
OB stars have revealed the presence of 
large-scale (quasi)cyclical structures, which
may be induced by instabilities generated by the star itself
\citep[see, e.g.,][]{kf98}. In addition, theory predicts the
existence of smaller-scale stochastic structure, caused by the
appearance of shocks in the 
wind flow, directly related to its intrinsic instability. In
simulations \citep[e.g.,][]{ro02}, irregular variations in both
density and velocity develop at a small
height above the photosphere, leading to the appearance of structure
already at a distance $r\la2\,R_{*}$ from the centre of the star,
where the wind has only reached about half its terminal velocity. The
variations steepen into shocks that decelerate and compress rarefied
gas, collecting most of the mass into a sequence of dense clumps
bounded by shocks, which definitely dominate the wind structure beyond 
$r\ga3\,R_{*}$ \citep{ro02} and perhaps much closer to the stellar
surface \citep{puls06}. 

The observational evidence for these clumps is, at present, 
indirect. Structured winds with numerous internal shocks are
considered necessary to explain X-ray emission from isolated
hot stars \citep{feld97}. A clumpy structure was also invoked to explain
the overall non-thermal radio emission from early-type stars, though
this is now debated \citep{loo06}. Recent model fits to optical and UV
spectra of B-type supergiants seem to require a substantial amount of
clumping \citep{pri05, crow06}. The observed stochastic variability on short
timescales of H$\alpha$ profiles in O-type stars is consistent with the
predictions of wind models with clumping \citep{markova}. In spite of
all this, as we have no direct observations of wind clumps, even their
most basic parameters (size, geometry, density) are unknown. Some
authors find compelling reasons to argue for geometrically large, very
massive clumps \citep{oski07}, but this is hotly debated. In any case,
the approximation of a smooth wind of constant density must be
considered unrealistic. 

In addition, the overall geometry of the wind is the subject of
debate. Radiation driven wind theory assumes a spherical geometry, but
other physical effects, such as stellar rotation and magnetic fields,
may cause important deviations. Some indirect evidence for the
presence of equatorially enhanced mass loss, perhaps associated to
magnetic fields, has been presented \citep{markova, uddoula08}. On the other
hand, the nebulae ejected by some massive stars are decidedly bipolar,
suggesting enhanced mass loss at the poles \citep{st07}.

\subsection{Accretion on to a compact object}

In the Bondi-Hoyle-Lyttleton approximation, particles follow ballistic
trajectories and are accreted without any net transfer of angular
momentum. However, from the very earliest numerical calculations, it
was clear that accretion flows should be more complicated. Initial
hydrodynamical simulations were carried out in two dimensions
\citep[e.g.,][]{mat87,tf89} or considering an axisymmetric
three-dimensional problem \citep[e.g.,][]{hunt71,saw89}, but many 3D
schemes have been developed \citep[e.g.,][]{mat91,nag05}. The
details of the flow depend on the thermodynamical properties of the
gas accreted and the size of the accretor (cf., for instance, Ruffert
1994; Ruffert \& Anzer 1995, and other papers in that series). The
accretion rate predicted by the Bondi-Hoyle-Lyttleton scheme turns out
to be a good approximation, but the accretion flow is much more
complex than assumed, especially as more realistic
physics are included \citep[e.g.,][]{tj93}. Bow shocks form and
complex structure develops around the accretor. In the case when the
mass donor fills an important fraction of its Roche lobe, as happens
in most SGXBs, the accretion flow is especially complicated
\citep{nag05}. 

When a density or velocity gradient is introduced in the accretion
flow, it becomes possible to accrete angular momentum. High resolution
2D simulations by \citet{mat87} found that the accretion flow does not
approach a steady state, but becomes unstable. The shock cone
oscillates from one side of the accretor to the other, allowing the
appearance of transient accretion discs. This instability, generally
known as ``flip-flop'' oscillation, has been studied by a large number
of authors, \citep[e.g.,][]{mat91,ba94}. The instability produces
fluctuations in the accretion rate and gives rise to stochastic
accretion of positive and negative angular momentum, leading to the
suggestion that it is the source of the variations seen in the pulse
evolution of SGXBs \citep{benen97,shi98}. \citet{ab95} argued that the
timescales and amplitudes of these 
fluctuations are too small to explain the spin variations, though
larger amplitudes were found in 2D simulations by \citet{shi98}.  If
the accretion flow is stable, the amount of angular momentum 
transferred to the neutron star is negligible. The more unstable the
accretion flow, the higher the transfer rate of angular momentum
\citep{mat91}. 

Unfortunately, the
results of simulations depend strongly on 
the procedure used. In general, the instability seems to appear when
the physical size of the accretor is very small, but full 3D
simulations are always much more stable than 2D simulations. Indeed,
several 3D simulations showed stable flows,  rising the possibility that the
flip-flop instability could be just a numerical artifact \citep{ruff99,
  kryu05}. A critical review of existing work is 
presented by \citet{fog05}, who suggest that the instability is most
likely physical in origin, related to the coupling of advected
perturbations to acoustic waves. Recent high-fidelity 2D numerical
simulations \citep{bp09} confirm that the flip-flop instability does not
require any gradient in the upstream flow to develop, and that it is
much stronger when the accretor is very small. If the accretor is
sufficiently small, the secular evolution is 
described by sudden jumps between states with counter-rotating
quasi-Keplerian accretion disks. 

In  addition to the intrinsic instability of the accretion flow, in a
real SGXB, the presence of the neutron star has an impact on the wind
dynamics \citep{ste88}. It may dynamically induce an enhancement of
the mass loss (leading to the formation of an accretion stream;
cf.~Blondin et al. 1991) and it may even induce the supergiant to pulsate
\citep{quaint03}. Also, if the neutron star is actively emitting X-rays, it
can impact on the wind by photo-ionising the heavy elements that drive
the outflow. In this way, it may slow down the wind, effectively
increasing the accretion radius (Blondin et 
al. 1991). Indeed, the photo-ionisation effect of the X-ray source is
readily visible in the ultraviolet spectra of SGXBs. Wind lines show
pronounced orbital variability \citep{kap93}. Far ultraviolet spectra
of the counterpart to 4U~1700$-$37 show important changes in the
ionisation structure of the wind with the binary phase, suggesting
that the X-ray source ionises much of the wind \citep{iping07}. In the optical
counterpart to Vela X-1, optical spectra reveal additional absorption
components on the wings of photospheric lines at inferior conjunction
of the neutron star \citep{kap94}. These features are attributed to
the presence of a photo-ionisation wake trailing the neutron star.

The complexity of the accretion environment can provide an explanation
to the short-term variability of the X-ray lightcurves of
SGXBs. Simulations by \citet{blond90} showed that the accretion wake
trailing the neutron star contains dense filaments (up to 100 times
denser than the undisturbed wind). Their accretion may produce abrupt
changes in X-ray luminosity.

\subsection{Magnetic field}

The neutron stars in SGXBs are young, and possess strong
($\sim10^{12}$~G) magnetic fields. As material approaches the neutron
star, it is captured by the magnetic field and deflected along the
magnetic field lines. Many situations are possible, depending on the
relative strengths of the magnetic field and the ram pressure of the
incoming material \citep{swr86,bozzo08}. If the ram pressure is low,
the incoming material may be stopped at the magnetosphere, and the neutron
star may then be in the propeller regime \citep{is75}. Accretion is
then told to be centrifugally inhibited. If the magnetic field is
strong enough to affect the flow beyond the accretion radius, there
will be magnetic inhibition of the accretion
\citep{swr86}. 

Though changes in the accretion state have been invoked to explain the
behaviour of Be/X-ray transients \citep[e.g.,][]{cam02}, magnetic
effects have not been traditionally
considered while explaining the long-term evolution of SGXBs. However,
the spin fluctuations (which are also seen in systems with an
accretion disc; e.g., Bildsten 1997) are highly suggestive of magnetic
interaction between the accretion 
disk and the stellar magnetosphere. \citet{ab95} argued that a disc or
torus could form around the magnetosphere and interact with it, acting
as a reservoir of mass 
and angular momentum. This interaction represents a possible way to
provide the neutron star with enough angular momentum to explain
random variations in pulse period, though \citet{ab95} estimate that
magnetic fields too high by an order of magnitude are needed for the
torque to be effective.

\section{Case study: GX~301$-$2}

Though Vela X-1 is probably the best studied SGXB, an important effort
has also been dedicated to understanding a very special system,
GX~301$-$2. This wind accreting X-ray pulsar differs from other SGXBs
in two very 
important aspects: first, its orbit is wider ($P_{{\rm orb}}=41.5$~d)
and more eccentric ($e=0.46$) than those of other systems; second, the
mass donor is a B1\,Ia$^{+}$ hypergiant, implying that it is larger
than the donors in other systems and has a denser, slower
wind. Because of these characteristics, the X-ray luminosity is
strongly modulated at the orbital period, The flux close to periastron
is $\sim 4$ times higher than at other phases, peaking 1--2~d before
periastron \citep[e.g.,][]{wat82,koh97}. The varying conditions along
the orbit make GX~301$-$2 a very good test of different accretion
models. 

GX~301$-$2 occasionally undergoes rapid spin-up episodes, lasting a
few weeks, which have been attributed to the formation of transient
accretion discs \citep{koh97}. In order to explain the existence of
strong periastron and weak apastron flares, \citet{pg01} postulated
the presence of circumstellar disc around the hypergiant, perhaps in
the form of a collimated, denser equatorial wind. Optical observations
of the mass donor, however, do not show any evidence for such a disc
\citep{kap06}. 

In fact, the size of the hypergiant is such that it may be very close
to filling its Roche lobe when the neutron star is at
periastron \citep{koh97}, allowing the formation of an accretion
stream. Some indications of the existence of such a stream are seen in
optical spectra of the companion \citep{kap06}. The average X-ray
lightcurve of GX~301$-$2 can be successfully 
fit by a model that considers accretion from a spherical wind and an
accretion stream \citep{lk08}. As a matter of fact, in this model,
most of the X-ray luminosity is due to accretion of dense material
from the stream, rather than direct wind accretion.

\section{An unexpected test on the model: supergiant fast X-ray
  transients}

Over the last few years, there has been a huge increase in the number
of HMXBs known. Many of these systems are believed to be wind
accretors \citep[e.g.,][]{wal06}. Several new SGXBs have been found,
most of them heavily obscured. Many other sources, however, have been
found to display very brief outbursts, with a 
rise timescale of tens of minutes and lasting only a few hours
\citep{sgue05}. They have been identified with OB supergiants:
 \citep[e.g., the prototype XTE~J1739$-$302;][]{smith06,neg06b},
 leading to the definition of a class of Supergiant Fast X-ray
 Transients \citep[SFXTs;][]{esa,smith06}. The
 distances to their counterparts imply typical $L_{{\rm
    X}}\sim10^{36}\:{\rm erg}\,{\rm s}^{-1}$ at the peak of the
outbursts. Outside outburst, they display rather lower luminosities
(between $\sim10^{33}$ and $\sim10^{35}\:{\rm erg}\,{\rm s}^{-1}$,
depending on the particular system; \citealt{sidoli08}).

As the number of these new systems
  increases, it is becoming clear that the separation between SFXTs
  and SGXBs is not well defined \citep{wzh07,neg08}. Some systems
  behave as SGXBs during part of their orbits (e.g.,
  SAX~J1818.6$-$1703, Zurita Heras \& Chaty 2008; IGR~J18483$-$0311,
  Rahoui \& Chaty 2008 ), and all show spectra and lightcurves typical
  of wind accretion.

Explaining the behaviour of these sources has become a major challenge
for our models of wind accretion. In some systems, variability is
obviously related to orbital motion. The orbital period of SAX~J1818.6$-$1703,
for example, is 30~d, and so its orbit is wider than those of SGXBs,
meaning that the neutron star is most of the time further away from
the donor than in the persistent systems. However, geometry alone does
not seem able to explain all the differences. The idea that the flares
are related to accretion of wind clumps was advanced by \citet{zand05}
and developed by \citet{wzh07}. \citet{neg08} speculated with the
possibility that the development of clumps made accretion less efficient
when the neutron star was relatively distant from the mass
donor. Alternatively, \citet{sidoli07} proposed that the outbursts
were associated with the crossing of a thin circumstellar disc
surrounding the supergiant. Other authors have suggested that the
reason of the outbursts must be sought in the interaction of the wind
with the magnetosphere \citep{greb07}, though \citet{bozzo08} calculated
that this was only plausible if the neutron stars in SFXTs had
magnetic fields two or three orders of magnitude stronger than those
in SGXBs.

\citet{wzh07} suggested that the X-ray lightcurves of SGXBs could be
tracing directly the matter density met by the neutron star. In this
way, the neutron star could be used as a probe of the stellar
wind. Unfortunately, detailed observations of outbursts from SFXTs do
not support the idea that each outburst can be identified with the
accretion of a wind clump. The outbursts last many hours (or even days) and are
generally multi-peaked. Many of them are also very asymmetric, with
fast rises and slow decays, casting doubt on the possibility that they
may be caused by direct accretion during passage through a disc. 

Moreover, intensive monitoring of SGXBs has also revealed new
phenomenology, which increases the similarity to SFXTs. Vela X-1 has
been observed to display ``giant'' flares, which resemble very closely
the outbursts of SFXTs, with flux increases of $\sim 10$ and a
timescale of a few hours \citep{krey08}. Conversely, it has also been
found to display ``off states'', during which the flux decreases
orders of magnitude and the pulse period is not
detectable. \citet{krey08} discuss the differences and similarities of
SGXBs and SFXTs in the context of a scenario in which wind clumping,
the highly structured accretion flow and the magnetic field all play a
role. 

More recently, the discovery that the SFXT IGR~J16479$-$4514 has an
orbital period of only 3.3~d \citep{jain09}, and that its outbursts
appear to be locked in phase \citep{bozzo09}, has posed a strong
challenge to all the models developed. Once more, the physics of wind
accretion reveals its complexity.

\section{Wind accretion in very high energy sources}

A handful of HMXBs produce $\gamma$-rays. To this day, two very
different models have been proposed to explain their behaviour (see
these proceedings). In both, the stellar winds play an important role,
but accretion is an important element in only one of them.

Of the $\gamma$-ray binaries, LS~2883/PSR B1259$-$63 is the only one
for which a model is universally accepted. PSR B1259$-$63 is a radio
pulsar in a very wide an eccentric orbit 
($P_{{\rm orb}}=3.4\:{\rm yr}$, $e=0.87$) around the Be star LS~2883. Timing of
the radio pulses provides an accurate orbital solution \citep{wex98}. 
As the neutron star approaches periastron, it interacts with the
disc around the Be star and radio pulses are quenched. The interaction
between the pulsar wind and the Be star outflow produces shocked material,
where particles are efficiently accelerated to relativistic velocities.
\citet{ta97} proposed a number of feasible mechanisms to explain the
production of high-energy photons. However, it must be noted that they
used a radial outflow model for the mass loss of Be stars
\citep{wat88}. This model has subsequently been ruled out by
observations, which strongly support a Keplerian disc around Be stars
\citep{pr03}. New modelling is necessary for this system (cf.~Naito,
these proceedings).

 The $\gamma$-ray binary LS I +61$^{\circ}$303 is similar to
 LS~2883/PSR B1259$-$63 in the sense that the mass donor is an early
 Be star and the orbit is highly eccentric \citep{cas05}. The orbital
 period is, however, much shorter (26 d). In the third  $\gamma$-ray
 binary, LS~5039, the situation is very different, as the mass donor
 is an O6.5\,V star in a close orbit \citep{clark01}. The wind
 structure is expected to be extremely different from those in the Be
 systems. However, it is worth noting that our knowledge of the winds
 of these systems is completely indirect. Only LS I +61$^{\circ}$303 was
 observed on a few occasions by IUE, but the spectra are of low
 quality \citep{howarth83}.

Moreover, with the small sample known, we have been unable to identify
the key ingredients necessary to have a $\gamma$-ray binary. For
example, the system SAX J0635+0533, which shows remarkable
similarities to PSR~B1259$-$63 and  LS I +61$^{\circ}$303 (a young
fast-spinning 
pulsar in an eccentric orbit around an early Be star; Cusumano et
al. 2000) is not a source of $\gamma$ rays \citep{aha06}.

An interesting possibility is opened by the proposed association of
the SFXT IGR J11215$-$5952 and the high-energy source EGR J1122$-$5946
\citep{sgue09}, though the error circle for the the EGRET source
is too large to consider this identification certain.

\section{Conclusions}

Supergiant X-ray binaries represent powerful laboratories to study the
physics of accretion from a wind. As our observations become more
extensive and sensitive, new phenomena are revealed, reflecting the
complexity of the physical situation. Increased computational
resources have provided very detailed models of wind accretion. These
models have shown that the accretion process is likely intrinsically
unstable, and perhaps dominated by stochastic processes. The most
important question that we need to understand is whether accretion
discs may form and angular momentum can be effectively transferred to
the accreting neutron star. Several observational facts strongly point
to a positive answer, but models still suggest that the effective
accretion of angular momentum is very difficult. New observations are
being taken and
more powerful models being developed. The next few years are likely to
see important developments in this field.


\acknowledgements 
This research has been funded by grants AYA2008-06166-C03-03 and
Consolider-GTC CSD-2006-00070 from the Spanish Ministerio de Ciencia e
Innovaci\'on (MICINN).



\end{document}